\documentclass[manuscript,screen]{acmart}
\AtBeginDocument{%
  }

\usepackage{xcolor}

\usepackage[shortcuts,acronym]{glossaries}
\makeglossaries

\begin{document}

\newacronym{AI}{AI}{Artifical Inteligence}

\newacronym{GenAI}{GenAI}{Generative Artificial Intelligence}

\newacronym{IT}{IT}{Information Technology}

\newacronym{LLM}{LLM}{Large Language Model}

\title{Unanticipated Effects of Generative AI on Expertise Pathways and Performance Perception in System Administration}
\subtitle{\today--author copy soliciting feedback}

\author{Rana Abou Khamis}
\orcid{0009-0008-7199-4109}
\affiliation{%
  \institution{School of Information Technology, Carleton University}
  \city{Ottawa}
  \country{Canada}}
\email{Rana.Aboukhamis@cmail.carleton.ca}

\author{Hala Assal}
\orcid{0000-0002-3306-0558}
\affiliation{%
  \institution{Department Systems and Computer Engineering, Carleton University}
  \city{Ottawa}
  \country{Canada}}
\email{Hala.Assal@cunet.carleton.ca}

\author{Ashraf Matrawy}
\orcid{0000-0001-9220-4630}
\affiliation{%
  \institution{School of Information Technology, Carleton University}
  \city{Ottawa}
  \country{Canada}}
\email{Ashraf.Matrawy@cunet.carleton.ca}

\renewcommand{\shortauthors}{Abou Khamis et al.}

\begin{abstract}
While industry discourse often emphasizes immediate productivity gains and frames \ac*{GenAI} primarily as a tool for automation \cite{brynjolfsson2025generative,bubeck2023paper, al2024enhancing, zhou2026impact}, the integration of \ac*{GenAI} into system administration may involve deeper shifts in professional practice that are not yet fully understood. 
Based on interviews with 14 IT professionals, we identify two major shifts. First, \ac*{GenAI} creates a compression of expertise pathways; with accelerating tasks, \ac*{GenAI}  reduces hands‑on learning, which traditionally built technical intuition and judgment. Second, \ac*{GenAI} introduces a shift in perceptions of performance, where AI‑assisted speed becomes the new norm, creating pressure, uneven comparisons across teams, and hidden labor around verification and correction. Together, these findings highlight how \ac*{GenAI} effects on technical environments extend beyond accelerated workflows, and can potentially lead to undesirable outcomes.

\end{abstract}

\maketitle

\section{Introduction}

Industry discourse and early research frequently highlight the dramatic productivity gains and automation potential of \ac*{GenAI}~\cite{brynjolfsson2025generative, bubeck2023paper, al2024enhancing}. Some even anticipated that it would replace entry-level jobs~\cite{woodruff2024knowledge}. Hence, professionals are increasingly turning to \ac*{GenAI} for its perceived benefits. Among them are system administrators (sysadmins)~\cite{markevych2023review,uyyala2023role}---professionals operating in high-stakes environments, where every decision carries organizational accountability, and time pressure is a constant.

Historically, sysadmin expertise is built through a developmental \emph{ladder}, where hands-on routine work, manual troubleshooting, and exposure to system failures over time serve as the training ground for developing technical intuition and professional judgment~\cite{goodall2009developing,dawson2018future,koschmann1987mind,ericsson1993role, armstrong2015development,hrebec2001survey}. As the \emph{``keepers of the machines''}~\cite{li2019keepers,limoncelli2007practice, barrett2004field,kandogan2012taming}, sysadmins' role demands deep technical judgment, diagnostic intuition, and the ability to make critical decisions in environments that are constantly growing in complexity. In many ways, the sysadmins are the final safeguard whose experience and situational awareness determine system stability during high-stakes situations.

With \ac*{GenAI} taking over many of the core expertise-building tasks, the impact of this technology in shaping the next generation of sysadmins comes into question. Thus far, we know remarkably little about how \ac*{GenAI} affects the everyday realities of the system administration field and sysadmins. In particular, little is known about how \ac*{GenAI} influences expertise development, verification practices, and perceptions of performance among system administrators.  We address this gap by exploring the effects of \ac*{GenAI} adoption beyond performance and automation.

We conducted an REB-approved semi‑structured interview study with 14 sysadmins and IT professionals who use \ac*{GenAI} in their daily work. Participants have between 2 and 15 years of professional experience. We explored how \ac*{GenAI} is integrated into system administration workflows, as well as sysadmins' perceptions thereof. We used Braun and Clarke's reflexive thematic analysis~\cite{braun2006using} to analyze interview transcripts.

We bring the community's attention to two unexpected and concerning themes that have emerged from our analysis:  
\begin{itemize}

\item \emph{The Compression of traditional expertise pathways:} \ac*{GenAI} allows users to skip many foundational, hands‑on tasks that once served as the training ground for technical judgment~\cite{goodall2009developing,dawson2018future}. This raises questions about how professional expertise develops when early learning steps are shortened or bypassed.

 \item \emph{Performance perception shift:} As \ac*{GenAI} accelerates work, fast output becomes the expected norm. Yet this visible speed often hides the substantial human effort required for validation, testing, and verification. This shift influences both organizational and self‑evaluation of productivity.
 \end{itemize}

\section{ Compression of Traditional Expertise Pathways}
System administration expertise has traditionally developed through a progression from routine fixes to complex architectural decisions, and is built through repetition, troubleshooting, and learning from mistakes~\cite{armstrong2015development,hrebec2001survey}. Our findings show that \ac*{GenAI} introduces a tension between accelerated task completion and deeper experiential learning that supports technical intuition and judgment.

\textbf{GenAI Accelerates Routine Work.}~ 
As expected, participants found \ac*{GenAI} tools to be useful in accelerating seemingly mundane tasks. They consistently described \ac*{GenAI} as highly effective, but mainlu for routine tasks such as scripting, troubleshooting, and documentation. P13 explained that \ac*{GenAI} mainly helps with \textit{``tweaking or making changes to existing scripts,''} whereas P5 observed that \textit{``hardly any new solutions [are] generated''} with \ac*{GenAI}. All in all, participants rely on \ac*{GenAI} to reduce the hands‑on effort required for repetitive operational work.

\textbf{GenAI Extends Expertise Boundaries.}~ 
Participants also described \ac*{GenAI} as lowering the barrier to more advanced tasks. P10 described himself as \textit{``not a programmer,''} yet he uses \ac*{GenAI} to build bots across multiple programming languages despite not being familiar with them. Similarly, P12 created Power BI dashboards in \textit{``half a day''}--work typically performed by specialized business analysts. Several participants described the experience as \textit{``like having a teacher right next to you.''} These examples suggest that practitioners can now readily achieve outcomes that once required formal training, mentorship, or years of accumulated experience.

\textbf{GenAI Relies on Human Expertise.}~Participants consistently emphasized that while \ac*{GenAI} helps them accelerate outcomes, it does not replace the need for human expertise  especially in complex, contextual, or high-stakes tasks. Several participants described \ac*{GenAI} as useful for brainstorming or as a starting point, yet it remains incapable of producing complete or reliable solutions. As P9 explained, \ac*{GenAI} \textit{``is not there yet to basically develop a solution.''} Others stressed that architectural and design decisions still require human reasoning and  that \textit{``there has to be always a human intervention''} [P11].  Participants also expressed the need for caution when using \ac*{GenAI}. For example, P10 conveyed his skepticism, \textit{``I would always assume that it's wrong... especially for critical things,''} and P5 described using \ac*{GenAI} only as a \textit{``guide''} that still requires \textit{``my own diligence.''} At the same time, participants recounts highlight a noticeable shift in how expertise is exercised. As \ac*{GenAI} handles more of the execution, sysadmins increasingly find themselves evaluating, correcting, and refining AI-generated output rather than devising solutions. As P13 put it, \ac*{GenAI} may complete \textit{''60 to 70\%''} of a task, but the essential \textit{``tweaking''} still requires expertise. P14 compared \ac*{GenAI} to an \textit{``autopilot''}, a tool that can manage routine functions but still needs \textit{``a human in the cockpit.''} 

\textbf{Changing Pathways to Expertise.}~
Together, the above observations illustrate how pathways to expertise are changing. While \ac*{GenAI} accelerates task execution and allows less-experienced sysadmins to produce work quickly, it simultaneously shortcuts expertise development by eliminating the traditional sequence of hands-on practice, mentorship, and iterative problem solving. P6 noted that \textit{``without having the fundamental understanding to tell if that answer is right or wrong,''} sysadmins may blindly trust \ac*{GenAI} outputs. Some participants have started noticing such effects with interns in their organizations who appear to lack understanding of foundational concepts. Thus, the shift towards accelerated workflows, output validation rather than problem solving, combined with reduced hands-on practice may have detrimental effects on the very sysadmin expertise necessary to validate \ac*{GenAI} output and to perform complex tasks. Our findings raise concerns about how sysadmin expertise will develop, especially for future sysadmins who may bypass the experiential foundations needed for contextual reasoning and professional judgment.

\section{ Performance Perception Shift} 

Across interviews, we found that not only did \ac*{GenAI} change workflows, but it also deeply changed how performance is perceived. As AI‑assisted speed becomes routine, organizations and practitioners begin to recalibrate the baseline and what is viewed as \emph{``normal''} productivity. Frustration and a sense of slowness are felt when work is completed manually. Interestingly, the human effort behind AI‑generated work remains invisible and is not accounted for when evaluating productivity.

\textbf{Acceleration as a New Performance Baseline.}  Participants reported a dramatic compression in task completion time. Tasks such as configuration, reporting, and analysis became \textit{``exponentially''} [P1] faster. These multi‑day tasks now take but a few minutes to complete. 
P14 explained \textit{``if I wanted to create a workflow... manipulating the data, cleaning it... that would take 3–4 days. But with GenAI .... get the code and steps in 15–30 minutes.''} \ac*{GenAI} has become indispensable for some participants. P12 noted, \textit{``I cannot see myself working without AI.''} As accelerated timelines become normalized, those who opt not to use \ac*{GenAI} and those conducting necessary manual work risk being seen as slow or inefficient.

\textbf{Hidden productivity costs.} Despite the appearance of speed, our analysis shows that participants performed substantial unseen labor to ensure the usability of \ac*{GenAI} outputs. In addition to the time spent verifying outputs, participants described the need to write multiple prompts and to refine their prompts repeatedly for the same task. For example, P14 explained that he must \textit{``explain the context repeatedly''} when working in complex environments. This iterative refinement process is essential but was seen by participants as \textit{``almost annoying''} [P10], and is largely invisible in the final analysis of productivity.

\textbf{Shifts in perception of own productivity.} Our data analysis also revealed a shift in how participants evaluated their own productivity. Even when doing work manually, participants evaluated their productivity against the accelerated baseline. P2 explained that returning to manual research felt \textit{``emotionally... bad''} and could be frustrating because of the stark efficiency gap---a task that takes the participant a full day to complete can be completed in 15--30 minutes when using \ac*{GenAI}.   
This frustration indicates that the perceived ''normal'' pace of work has shifted. For some, the tool has transitioned from a convenience to an essential component of their professional effectiveness. P12 noted that \ac*{GenAI} \textit{``did boost my productivity and made it more efficient in very unexpected ways [...] I cannot see myself working without AI''}. These accounts demonstrate that \ac*{GenAI} is inadvertently shifting the baseline of what sysadmins perceive as a competent and productive pace of work.

\textbf{Team and Organizational Comparison. } Participants also described emerging tensions when access to \ac*{GenAI} was uneven across teams. For example, some teams, but not others, may be restricted from using \ac*{GenAI} due to privacy or security considerations. This can contribute to what several participants referred to as a \textit{``two‑speed culture;''} the team not using \ac*{GenAI} may appear less productive despite performing equally demanding work. P1 illustrated this clearly, 
\textit{``You have team member A that can use GenAI [and] team member B cannot use GenAI. So you give team member A ten tasks, and he finishes it very quickly. Team member B cannot use GenAI, and he is still struggling and spending weeks and even months on his own task. So it is putting undue pressure on team member B, whose own work is slow, and it seems as if he is not working as effectively and as hard as the person using GenAI''}~[P1]. On the organizational level, participants observed that once AI-supported workflows are introduced, the accelerated speed quickly becomes normalized. For example, tasks traditionally completed in months are expected to be delivered in a fraction of that time, reinforcing the perception that AI-assisted speed is the default baseline. P13 indicated that some project delivery times fell from \textit{``2 months''} to \textit{``2 weeks.''} These dynamics suggest the emergence of a two-tier work environment in which visible output becomes a primary indicator of performance, potentially obscuring differences in available tools, constraints, and working conditions.

\section{Discussion}

Our findings point to two connected shifts in how \ac*{GenAI} is reshaping system administration work. Together, they show that \ac*{GenAI} is not just accelerating tasks, but also altering how expertise develops and how performance is evaluated within system administration practice.

\textbf{Compression of the Experience Ladder. } \ac*{GenAI} is changing how people reach technical outcomes. It allows sysadmins, especially novices, to readily complete tasks without the need to go through the traditional route of years of hands‑on practice, mentorship, or deep independent research. The work is accomplished, but the pathway is different. This raises a central question: \textit{If early foundational learning steps are increasingly bypassed, how will future sysadmins develop the experiential grounding needed for complex, high‑stakes decisions?} \ac*{GenAI} lacks the ability to complete complex tasks, and its output requires validation, both of which require expertise. However, \ac*{GenAI} is offloading the very tasks through which this expertise is developed. The question remains on whether and how future generations of sysadmins will be able to develop the expertise necessary to complete complex tasks.

\textbf{Expertise Shifting From Production to Evaluation. } Across interviews, participants stressed that \ac*{GenAI} outputs cannot be accepted without scrutiny. They tested scripts, checked documentation, iterated prompts, and validated recommendations. The manual labor has not disappeared, but it has shifted from producing solutions to validating them. This raises a second question: \textit{If expertise is increasingly exercised through validation, how will future practitioners gain the depth of knowledge required to judge AI‑generated output?} Such depth and understanding of nuances behind decisions is typically built through years of hands-on troubleshooting experiences and exposure to failures. With \ac*{GenAI} essentially eliminating this, will future sysadmins continue to have the same level of efficiency in judging \ac*{AI} outputs? 

\textbf{Invisible Labor and Recalibrated Performance Expectations.}  Our analysis revealed a growing gap between the productivity of AI‑assisted work and the labor it actually requires. While organizations may see accelerated results, practitioners experience a longer, invisible workload of prompting, correcting, testing, refining, and validating. This hidden labor of \emph{babysitting} \ac*{GenAI} remains essential for producing safe and reliable results, yet it is often overlooked when performance is judged without taking full effort into consideration. As AI-assisted speed becomes the norm, performance expectations shift. Sysadmins are pressured to work faster, and frustrations arise when \ac*{GenAI} outputs require substantial correction. A two‑tier culture is created when teams with and without \ac*{GenAI} access are compared. Even on the individual level, our analysis shows that sysadmins are evaluating their own productivity based on the new \ac*{GenAI}-enabled baseline. They now perceive their performance prior to \ac*{GenAI} as inadequate, and the very thought of reverting back to completing tasks without \ac*{GenAI} is inconceivable.

\section {Conclusion}
We identified two unanticipated socio-technical effects of \ac*{GenAI} in system administration. First, by enabling sysadmins to achieve high-level tasks without going through the usual process of repeated practice, troubleshooting, and learning from errors, \ac*{GenAI} appears to compress traditional expertise pathways. While it undeniably facilitates technical work, it is  potentially creating a long-term \emph{``learning debt''}. Second, \ac*{GenAI} is creating a shift in performance perception, setting new baseline expectations for productivity, without accounting for the necessary and intensive manual verification and validation work performed behind the scenes. 
Although our findings are within the context of system administration workflows, we expect that these unanticipated effects also exist in other technical fields. We hope this paper draws the community's attention to an overlooked area that warrants further investigation.

\bibliographystyle{ACM-Reference-Format}
\bibliography{bibFile}

\end{document}